\documentstyle[twoside,fleqn,espcrc2,psfig]{article}

\newcommand{\ltsima} {$\; \buildrel < \over \sim \;$}
\newcommand{\gtsima} {$\; \buildrel > \over \sim \;$}
\newcommand{\lta} {\lower.5ex\hbox{\ltsima}}
\newcommand{\gta} {\lower.5ex\hbox{\gtsima}}

\newcommand{\nodata} {\raise.5ex\hbox{....}}

\newcommand{\AmS}{{\protect\the\textfont2
  A\kern-.1667em\lower.5ex\hbox{M}\kern-.125emS}}

\title{ {\it Beppo}SAX Observations of the TeV Blazar Mkn~421}

\author{
G. Fossati\address{SISSA/ISAS, via Beirut 4, 34014 Trieste Italy},  
L. Chiappetti\address{Istituto per ricerche in Fisica Cosmica e Tecnologie 
		      Relative, CNR, via Bassini 15, 20133 Milano},  
A. Celotti$^{\rm a}$, 
G. Ghisellini\address{Osservatorio Astronomico di Brera, via Bianchi 46, 
		      22055 Merate (LC), Italy},        
L. Maraschi\address{Osservatorio Astronomico di Brera, via Brera 28, 
		    Milano, Italy}                  
G. Tagliaferri$^{\rm c}$, 
E.G. Tanzi$^{\rm b}$, 
A. Treves\address{Universit\`a di Milano, Istituto di Scienze, II Facolt\`a
                  di Scienze, Via Lucini 3, I-22100, Como}, 
L. Bassani\address{ITESRE/CNR, via Gobetti 101, 40129 Bologna, Italy}, 
M. Cappi$^{\rm f}$, 
A. Comastri\address{Osservatorio Astronomico di Bologna, 
                    via Zamboni 33, 40126 Bologna, Italy}, 
F. Frontera$^{\rm f}$, 
S. Giarrusso\address{IFCAI/CNR, Via Ugo La Malfa 153, 90146 Palermo, Italy} 
P. Grandi\address{Istituto di Astrofisica Spaziale, CNR, via Fosso del Cavaliere,                   00133 Roma, Italy},  
S. Molendi$^{\rm b}$, 
G. Palumbo$^{\rm g}$, 
C. Perola\address{Universit\~a degli Studi di Roma, Italy},  
E. Pian$^{\rm f}$, 
M. Salvati\address{Osservatorio Astrofisico di Arcetri, Largo E.Fermi 5, 
                   50125 Firenze, Italy},   
C. Raiteri\address{Osservatorio Astronomico di Torino, 
                   Pino Torinese, Italy}, 
M. Villata$^{\rm l}$, 
C.M. Urry\address{Space Telescope Science Institute, 3700 San Martin Dr., 
                  Baltimore, MD 21218}
}

\begin{document}

\begin{abstract}
The blazar Mkn 421 has been observed, as part of the  AO1 Core Program,
five times from 2 to 7 May 1997. 
In the LECS+MECS energy band the spectrum shows convex curvature, well 
represented by a broken power--law. 
Flux variability (more than a factor 2) has been detected over
the entire 0.1--10 keV range, accompanying which the spectrum steepens 
with the decrease in intensity.   
Mkn 421 has also been detected with the PDS instrument. Our preliminary
analysis indicates that the PDS spectrum lies significantly above the
extrapolation from the MECS, suggesting a contribution from a flatter high
energy component. 
\end{abstract}

\maketitle

\section{Introduction}

Mkn 421 is one of the best known and studied BL Lac objects. 
It shows optical polarization, flat radio spectrum, and large 
variability, characteristics of the blazar class. 
It is bright in X--rays, where it shows prominent flares 
accompanied by significant spectral changes \cite{takahashi}.

Among GeV--emitting blazars, Mkn 421 is unique in being the 
first object in which the $\gamma$--ray emission is detected to extend 
up to TeV energies (\cite{punch,krennrich}), at a level allowing detailed 
spectral and variability studies in the broadest available range 
of frequencies.
Its GeV (EGRET) $\gamma$--ray emission connects smoothly with the
E$>$0.5 TeV spectrum (e.g.~\cite{macomb}).
On the contrary the X--ray spectrum, up to 10 keV, did not show any 
hint of the onset of the inverse Compton component responsible for GeV
and TeV emission.
It was then of great interest to take advantage of the full capabilities
of {\it Beppo}SAX, enabling a spectral coverage up to $\gta$ 200 keV, to
look for it.
At the same time the fact that in the X--ray band we are observing the
emission from the highest energy tail of the emitting--particle
distribution could provide important clues on particle acceleration and
cooling mechanisms.

The Mkn~421 observations discussed here are part of the AO1 Core Program
dedicated to bright blazars. 
The data reduction presented here has been generally
performed with software released {\it before} September 1997.
Data reduction with the updated software for all the
on board instruments, and a more detailed analysis, will be presented in
forthcoming paper with all appropriate references. 

\begin{figure}[t]
\vspace{9pt}
\psfig{file=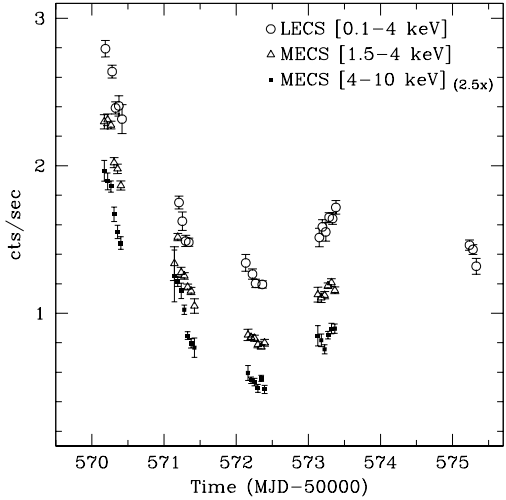,width=7.5truecm,rheight=5.3truecm}
\caption{\small\sf Rebinned light curve (4000~s bins) of {\it Beppo}SAX 
data, LECS, and MECS divided in two energy bands [1.5 -- 4], [4 -- 10] keV.}   
\label{fig:light_curve}
\end{figure}

\begin{table*}[t]
\setlength{\tabcolsep}{0.7pc}
\newlength{\digitwidth} \settowidth{\digitwidth}{\rm 0}
\catcode`?=\active \def?{\kern\digitwidth}
\caption{{\it Beppo}SAX Observations Log}
\label{tab:log}
\begin{tabular*}{\textwidth}{@{}l@{\extracolsep{\fill}}ccccc}
\hline
{Pointing} &            
{LECS } &            
{LECS } &            
{MECS } &            
{MECS } &            
{MECS } \\
{Start Date} & 
{exp. time} &            
{[0.1--4 keV]} &            
{exp. time} &            
{[1.5--10 keV]} & 
{[$<$4/$>$4 keV]} \\
{ } &
{(ksec)} &
{(cts/s)} &
{(ksec)} &
{(cts/s)} &
{(cts/s)} \\
\hline

{ 2/V/1997 @ 04:10 } & 
{  4.4 } & {$ 2.806 \pm 0.020 $} & { 11.4 } & {$ 2.862 \pm 0.016 $} & 
{ (2.14/0.72) } \\
{ 3/V/1997 @ 03:24 } & 
{  4.3 } & {$ 1.748 \pm 0.020 $} & { 11.7 } & {$ 1.694 \pm 0.012 $} & 
{ (1.28/0.41) } \\
{ 4/V/1997 @ 03:25 } & 
{  4.9 } & {$ 1.362 \pm 0.017 $} & { 12.2 } & {$ 1.027 \pm 0.009 $} & 
{ (0.81/0.22) } \\
{ 5/V/1997 @ 03:32 } & 
{  4.9 } & {$ 1.823 \pm 0.029 $} & { 11.9 } & {$ 1.522 \pm 0.012 $} & 
{ (1.52/0.35) } \\
{ 7/V/1997 @ 04:47 } & 
{  6.0 } & {$ 1.612 \pm 0.016 $} & { \nodata } & { \nodata } & 
{ \nodata } \\
\hline
\end{tabular*}
\end{table*}

\section{Observations and Data reduction}

For an exhaustive description of the Italian/Dutch {\it Beppo}SAX  mission we
refer to \cite{boella97}. 
The observations of Mkn~421 were performed as part of the {\it Beppo}SAX
AO1 Core Program between May 2$^{\rm nd}$ and May 7$^{\rm th}$ 1997. 
In this period Mkn~421 has been observed 5 times, for a total effective
exposure of 47.2 ksec in the MECS and 24.5 ksec in the LECS.  
A journal of observations is given in Table~\ref{tab:log}. 
MECS data are not available for May 7$^{\rm th}$ because of the failure
of the detector unit 1 on May 6$^{\rm th}$..
The present analysis is based on the {\tt SAXDAS} linearized 
event files for the LECS and the three MECS experiments, together with
appropriate background event files, and PDS 
as produced at the {\it Beppo}SAX Science Data Center ({\tt rev0.0}).  

\section{Data analysis and results}

Light curves and spectra for the LECS and the three MECS have been
accumulated using the {\tt SAXselect} tool. Events were extracted from
within a radius of 8' and 4' for the LECS and MECS images, respectively.  
These selected events were then used to construct the
light curve and to accumulate energy spectra for each pointing.  
LECS data have been considered only in the range [0.12--4] keV due to
calibration problems at higher energies (Guainazzi 1997, private
comm.).  
Spectral analysis has been performed with the{ \tt XSPEC 9.0}
package, using response matrices released on December 1996.

\begin{figure}[t]
\vspace{9pt}
\psfig{file=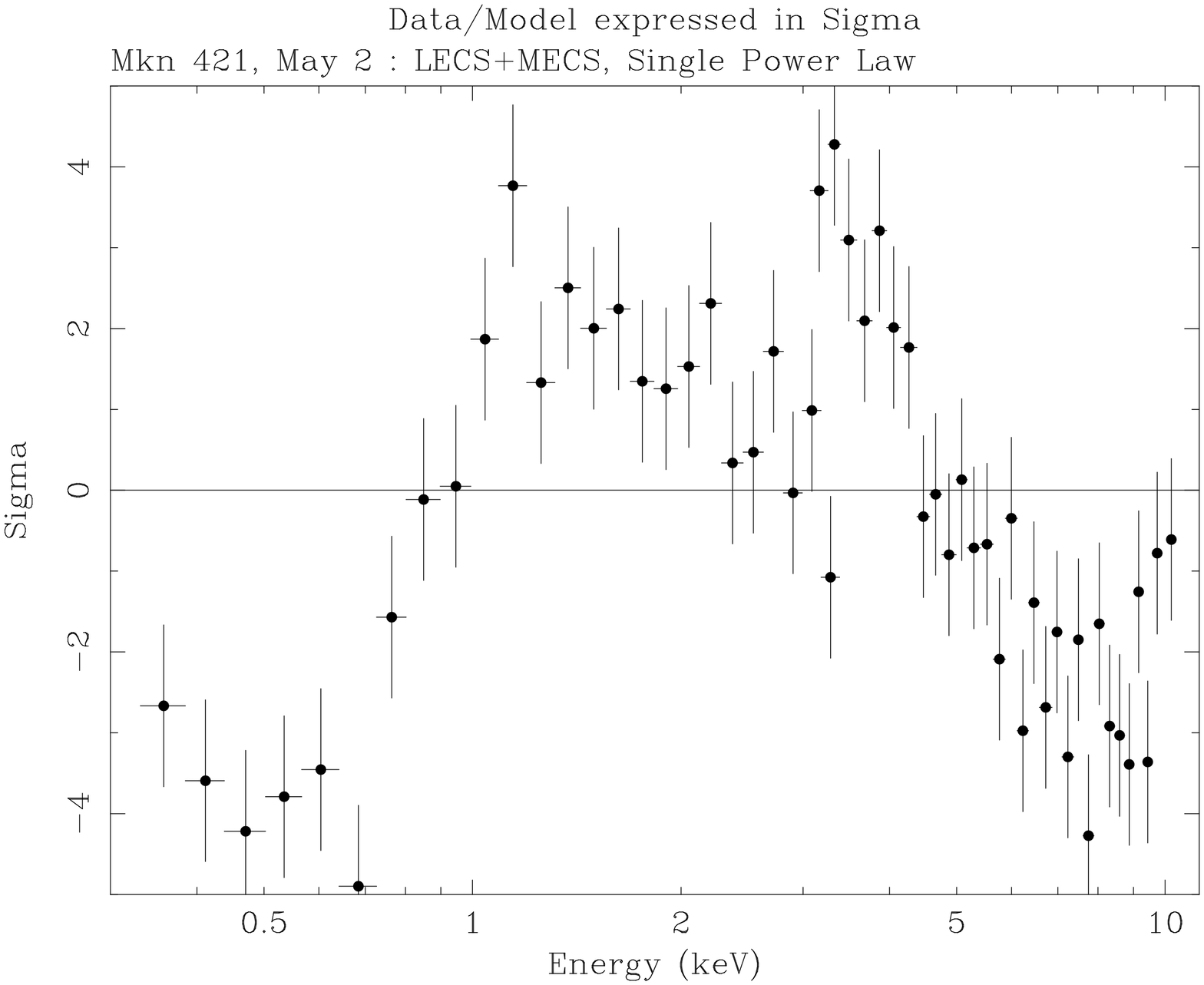,width=7.5truecm,height=3.0truecm,rheight=3.5truecm,angle=270}
\psfig{file=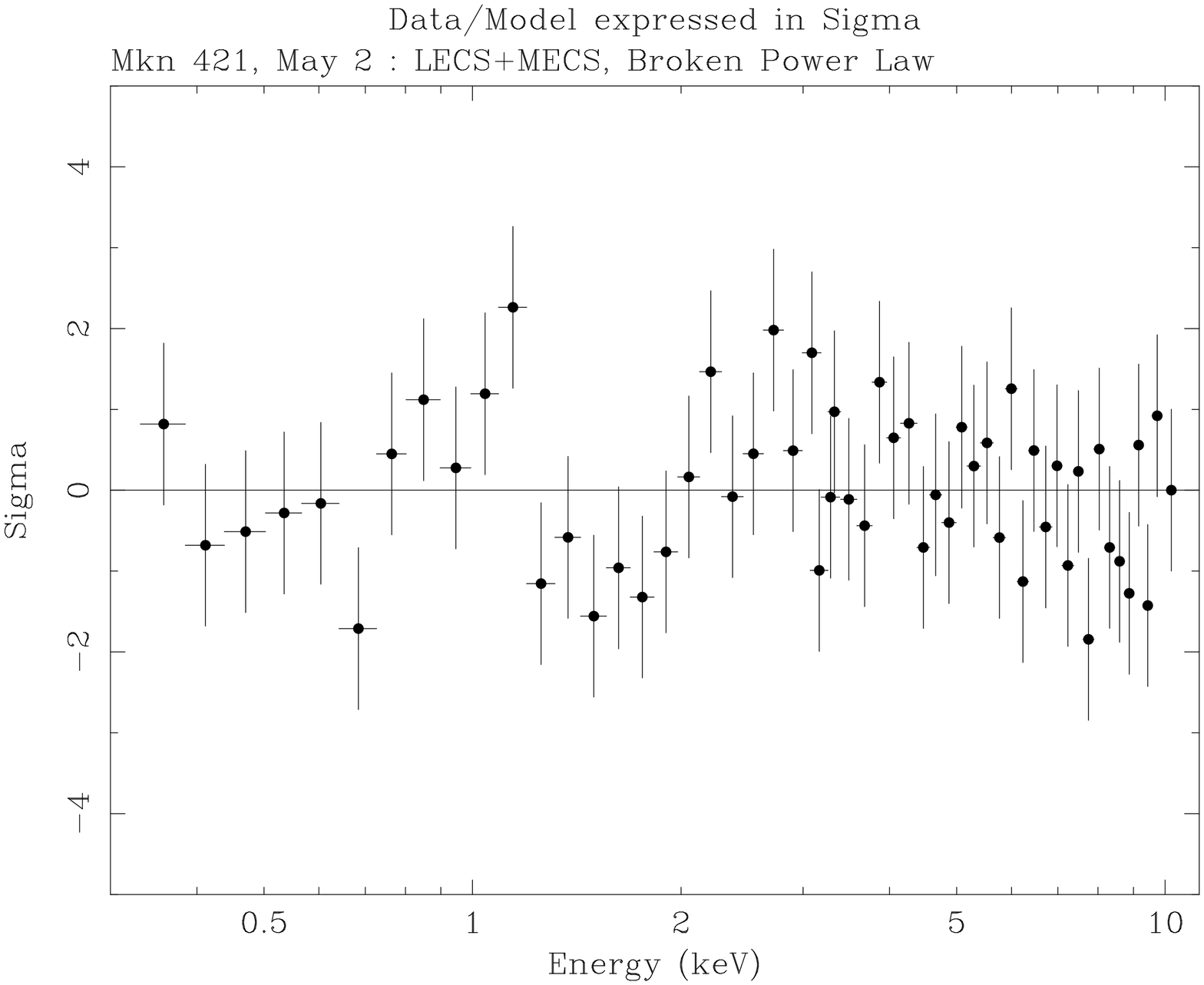,width=7.5truecm,angle=270,height=3.0truecm,rheight=2.2truecm}
\caption{\small\sf (a) 
Deviation of LECS+MECS data from the best fit single power law model 
(with Galactic absorption), for May 2, expressed in units of {\it sigma}.
(b) Deviation of LECS+MECS data from the best fit broken power law model.} 
\label{fig:data_fit}
\end{figure}

\begin{table*}[htb]
\setlength{\tabcolsep}{0.5pc}
\settowidth{\digitwidth}{\rm 0}
\catcode`?=\active \def?{\kern\digitwidth}
\caption{Single Power Law Spectral Fit$^{a,b}$}
\label{tab:fit}
\begin{tabular*}{\textwidth}{@{}llccccc}
\hline
{Date} & 
{Instrument} &           
{Model} & 
{$\alpha_1$} &
{E$_{\rm break}$} &
{$\alpha_2$} &
{$\chi^{2}_{\nu}$ ($d.o.f$)} \\
{} &           
{} &           
{} &           
{} &
{(keV)} &
{} &
{} \\
\hline
\underbar{\bf May 2} & {LECS$^c$} & { P.L. } & { 1.19$^{+0.02}_{-0.02}$ } & { --- } & { --- } & { 3.99 (32)} \\ 

& {LECS$^c$} & {broken P.L. } &  { 1.04$^{+0.04}_{-0.04}$ } & { 1.20$^{+0.19}_{-0.14}$} & { 1.49$^{+0.09}_{-0.07}$ } & { 0.94 (30)} \\ 

& {MECS} & { P.L. } & { 1.73$^{+0.02}_{-0.02}$ } & { --- } & { --- } & { 1.25 (54)} \\

& {MECS} & { broken P.L. } & { 1.65$^{+0.09}_{-0.06}$ } & { 4.24$^{+2.15}_{-0.71}$} & { 1.89$^{+0.37}_{-0.09}$ } & { 1.02 (52)} \\

& {LECS+MECS$^{c,d}$} & { broken P.L. } & { 1.05$^{+0.13}_{-0.13}$ } & { 1.51$^{+0.43}_{-0.23}$} & { 1.73$^{+0.07}_{-0.05}$ } & { 1.27 (48)} \\

\underbar{\bf May 4} & {LECS$^c$} & { P.L. } & { 1.41$^{+0.02}_{-0.02}$ } & { --- } & { --- } & { 2.82 (32)} \\

& {LECS$^c$} & {broken P.L. } & { 1.30$^{+0.03}_{-0.03}$ } & { 1.13$^{+0.15}_{-0.17}$} & { 1.68$^{+0.09}_{-0.07}$ } & { 1.70 (30)} \\

& {MECS} & { P.L. } & { 2.00$^{+0.02}_{-0.03}$ } & { --- } & { --- } & { 1.00 (54)} \\ 

& {MECS} & { broken P.L. } & { 1.67$^{+0.19}_{-0.41}$ } & { 2.93$^{+0.58}_{-0.54}$} & { 2.15$^{+0.11}_{-0.10}$ } & { 0.74 (52)} \\

& {LECS+MECS$^{c,d}$} & { broken P.L. } & { 1.35$^{+0.05}_{-0.04}$ } & { 3.33$^{+0.34}_{-0.26}$} & { 2.03$^{+0.13}_{-0.12}$ } & { 2.94 (56)} \\
\hline
\multicolumn{7}{@{}p{160mm}}{
$^{(a)}$ Quoted errors are at 1 $\sigma$, computed for value of
$\triangle\chi^2$ appropriate for the number of interesting parameters; 
$^{(b)}$ The hydrogen equivalent column density
n$_H$ has been fixed at the Galactic value. Adopted value is 1.61
$\times$ 10$^{20}$cm$^{-2}$, from Lockman
and Savage 1995;  
$^{(c)}$ LECS data in the restricted range 0.3--4. keV;  
$^{(d)}$ MECS data in the restricted range between 3--10 keV.}  
\end{tabular*}
\end{table*}

\noindent
\underbar{\bf Temporal analysis} $\bullet$ 
the LECS and MECS light curves during the 7 days spanned by {\it Beppo}SAX
pointings are shown in Fig.~\ref{fig:light_curve}.

MECS data have been divided in two energy bands, below and
above 4 keV, in order to better show the different variability amplitude at
different energies. 
There is evidence of a significant spectral variability between the 
May 2$^{\rm nd}$ and May 4$^{\rm th}$ states.
The LECS count rate changes of a factor $\sim$ 2.2, while in the
higher MECS band (E $>$ 4.0 keV) the change is of a factor $\sim$ 4. 

\begin{figure}[bt]
\vspace{9pt}
\psfig{file=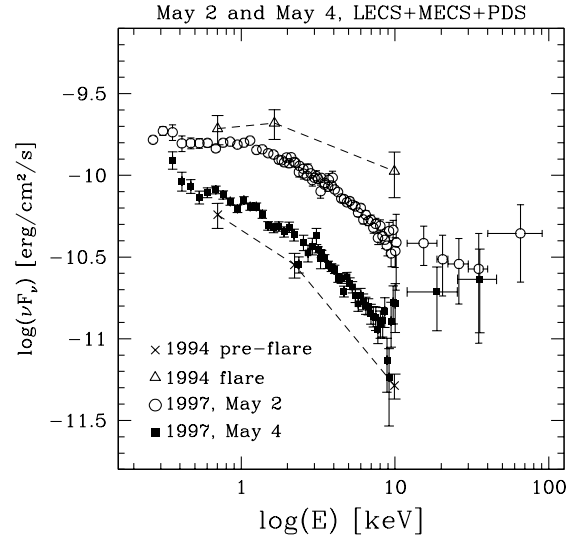,width=8.1truecm,rheight=5.8truecm}
\caption{\small\sf May 2$^{\rm nd}$ and 4$^{\rm th}$ 0.3--90 keV 
{\it reconstructed Beppo}SAX spectra, in $\nu F_\nu$ representation.
For reference we plot also the ASCA data corresponding to the 1994 pre--flare
and TeV flare states \protect\cite{macomb}, connected with dashed
lines.}
\label{fig:deconvolved}
\end{figure}

\noindent
\underbar{\bf Spectral analysis} 
\underbar{\it LECS, MECS} $\bullet$ the source was quite 
bright (see Fig.~\ref{fig:deconvolved} for a comparison) and consequently
the data are of good quality enabling us to extract spectra from each
observation. 
A word of caution is necessary about the presence of localized
features probably related with calibration problems yet to be solved (e.g.
$\sim$ 2 keV where there are Gold features due to the optics), 
At the present stage we only performed a preliminary analysis aimed at 
studying the continuum properties and their temporal variability.
Moreover, here we only concentrate on analysis and results concerning the
highest and lowest states observed by {\it Beppo}SAX, on May 2$^{\rm nd}$
and May 4$^{\rm th}$ respectively.

The resulting accumulated LECS and MECS data have been fitted with a
variety of combination of single and broken power law models to better 
describe the marked spectral curvature.
The main results of spectral fits are reported in Table~\ref{tab:fit},
and the ``reconstructed" spectra are shown Fig.~\ref{fig:deconvolved}.
Fig.~\ref{fig:data_fit}a shows the deviation of the data (LECS+MECS,
May 2$^{\rm nd}$) from a single power law model with Galactic absorbing
column density (N$_{\rm H}$ = 1.61 $\times 10^{20}$ cm$^{-2}$ \cite{lockman}).
It is clear that this model does not provide a satisfactory fit to the data 
($\chi^2_{\nu} = 4.25$ for 48 $d.o.f.$) which require a flatter
component below $\sim$ 1 keV, and steeper one beyond $\sim$ 3 keV.
Fig.~\ref{fig:data_fit}b shows the deviations with respect to
a broken power law model giving a marginally acceptable fit 
($\chi^2_{\nu} = 1.27$, see Table~\ref{tab:fit}, P$_{\chi^2,\nu} <$ 10 \%).
Apart from localized features it is still noticeable a steepening trend
above $\sim$ 4 keV, although consistent with the adopted model.
As can be seen in Table~\ref{tab:fit} the broken power law model gives a
better fit both in LECS and MECS energy ranges. 
LECS data, even though in the restricted interval 0.1--4.0 keV are not
consistent with a unique spectral index.
Values of $\alpha_1$ and $\alpha_2$ (LECS $\rightarrow$ MECS) clearly 
trace the continuous spectral steepening of the 0.1--10 keV spectrum.
Both May 2$^{\rm nd}$ and 4$^{\rm th}$ 0.3 -- 10 keV spectra
show a $\triangle \alpha \simeq 0.85 \pm 0.05$ (see Table~\ref{tab:fit}
LECS $\rightarrow$ MECS broken P.L.). 
On the other hand there is a marked spectral change accompanying the flux
change, in the sense of a steepening in the lower state, for E $< 10$ keV
($\triangle \alpha_{\rm LECS,low} \simeq 0.26$, $\triangle \alpha_{\rm
MECS,high} \simeq 0.26$).

\noindent 
\underbar{\it PDS} $\circ$ The source was weakly detected in the PDS 
at E $>$ 15 keV, although with low
significance ($\gta$ 2--3 $\sigma$), on May~2$^{\rm nd}$ up to 90 keV, 
on May~4$^{\rm th}$ only up to 50 keV.
Spectral analysis turns out to be delicate and is in progress. 
Results about PDS data are thus very preliminary.
PDS data are in excess with respect ot the extrapolation of the 
very steep ($\alpha = 1.89 - 2.15$) $>$4 keV spectrum, suggesting a 
contribution from a harder spectral component (see
Fig.~\ref{fig:deconvolved}).
Results will be presented in a forthcoming paper. 

\section{Conclusions}

{\it Beppo}SAX dat aof Mkn 421 shows interesting variability both in flux
and in spectral shape, with a marked softening corresponding to decreasing
brightness. 
This kind of spectral variability behavior is well known in the X--ray 
band for sources of the class of Mkn~421, the so called
High-Frequency-Peaked BL Lacs (HBL). 
In general in blazars at energies just above the
synchrotron peak it the relationship {\it harder-when-brighter} holds
and is generally interpreted in terms of injection of fresh electrons in
the highest energy end of a single population.

On May 2$^{\rm nd}$ the peak of the synchrotron component could possibly
fall in/or just below the LECS band, while in the lower state of 
May 4$^{\rm th}$ this is not longer true.

Moreover, preliminary analysis of PDS data suggests the presence of  
a deviation from the continuously downward curvature for E $>
10$ keV, possibly being the signature of the onset of a different harder
spectral component.

Further temporal and spectral analysis, also with comparison of
multifrequency data and comparison with model predictions is in progress.

\end{document}